# Inverse-designed non-volatile phase change varifocal metalens at the edge of the visible spectrum

SUBRAHMANYAM S. G. MANTHA[1,2], HARISH N. S. KRISHNAMOORTHY[1*]

*Tata Institute of Fundamental Research Hyderabad, Hyderabad 500046, India*
2. *Birla Institute of Technology and Science Pilani, KK Birla Goa Campus, Sancoale, Goa 403726, India*
*[*harishk@tifrh.res.in](mailto:harishk@tifrh.res.in)*

**Abstract:** Reconfigurable metalenses capable of large focal length tuning, fast response times, and high focusing efficiency while maintaining diffraction-limited operation are highly desirable for next-generation adaptive imaging systems. Phase change chalcogenides provide a promising platform for such devices by exploiting the reversible amorphous-to-crystalline transition to achieve non-volatile tuning with relatively fast switching. However, extending these approaches towards the visible spectrum is challenging because of the reduced meta-atom dimensions, stringent phase coverage requirements, intrinsic material absorption and the need to simultaneously preserve focusing efficiency and image quality across multiple material states. Here, we present a dynamically tunable metalens based on $Sb_2S_3$ operating at the edge of the visible spectrum. The design framework combines finite element computations with a genetic algorithm-based inverse design approach to achieve robust phase control in both amorphous and crystalline states. The resulting metalens shows near-diffraction-limited performance with a focal length tunability of 33%, Strehl ratios of 0.80 and 0.77 in the amorphous and crystalline states, respectively, and focusing efficiencies of approximately 40% which are among the highest reported for metalenses designed entirely within an active phase-change chalcogenide medium.

## 1. Introduction

Subwavelength material structuring enables precise control over light–matter interactions, allowing independent manipulation of optical phase, amplitude, and polarization, and thereby enabling advanced functionalities in photonic components such as sources, detectors, imaging systems, and sensors [1–5]. Among such innovations, metalenses – planar optical elements composed of spatially varying nanoscale structures have demonstrated precise wavefront engineering through spatial phase modulation [6–11]. Their compact form factor facilitates miniaturization of optical systems significantly, with demonstrated applications in endoscopic imaging, astronomical instrumentation, augmented reality displays, and holographic projection systems [12–15]. In addition, metalenses have been employed for ultracompact microscopy [16], scalable optical addressing of qubits in ion trap quantum processors [17], as well as for manipulating the emission and spatial mode control of single-photon sources, essential for photonic quantum computing architectures [18–20]. Metalens-based platforms have also been utilized to perform programmable quantum operations, including demonstration of Grover's quantum search algorithm [21]. A metalens achieves focusing by reproducing the lens' quadratic phase profile through engineered arrays of nanoscale scatterers. Fabricated via electron beam or focused ion beam lithography, metalenses are typically static – once realized, their optical properties such as operational spectral range, optical zoom and focal length are fixed. However, dynamic control of these properties is critical for advanced imaging systems, including those used in adaptive optics, biomedical imaging, and wearable vision technologies. Reconfigurability in optical properties of metalenses across the visible and infrared region has been explored via mechanical means [22,23], electrically activated thermo-optic effects [24], electro-optic effects [25], by employing liquid crystals [26], phase change materials [27–29], and also by optical means [30,31]. While these approaches enable tunability, most demonstrated systems operate in the infrared region. In the visible region, previously reported works either use designs where the metalens is itself made of a static medium or employ thermally/mechanically tunable material platforms which inherently restrict switching speeds [22,24–26,28].

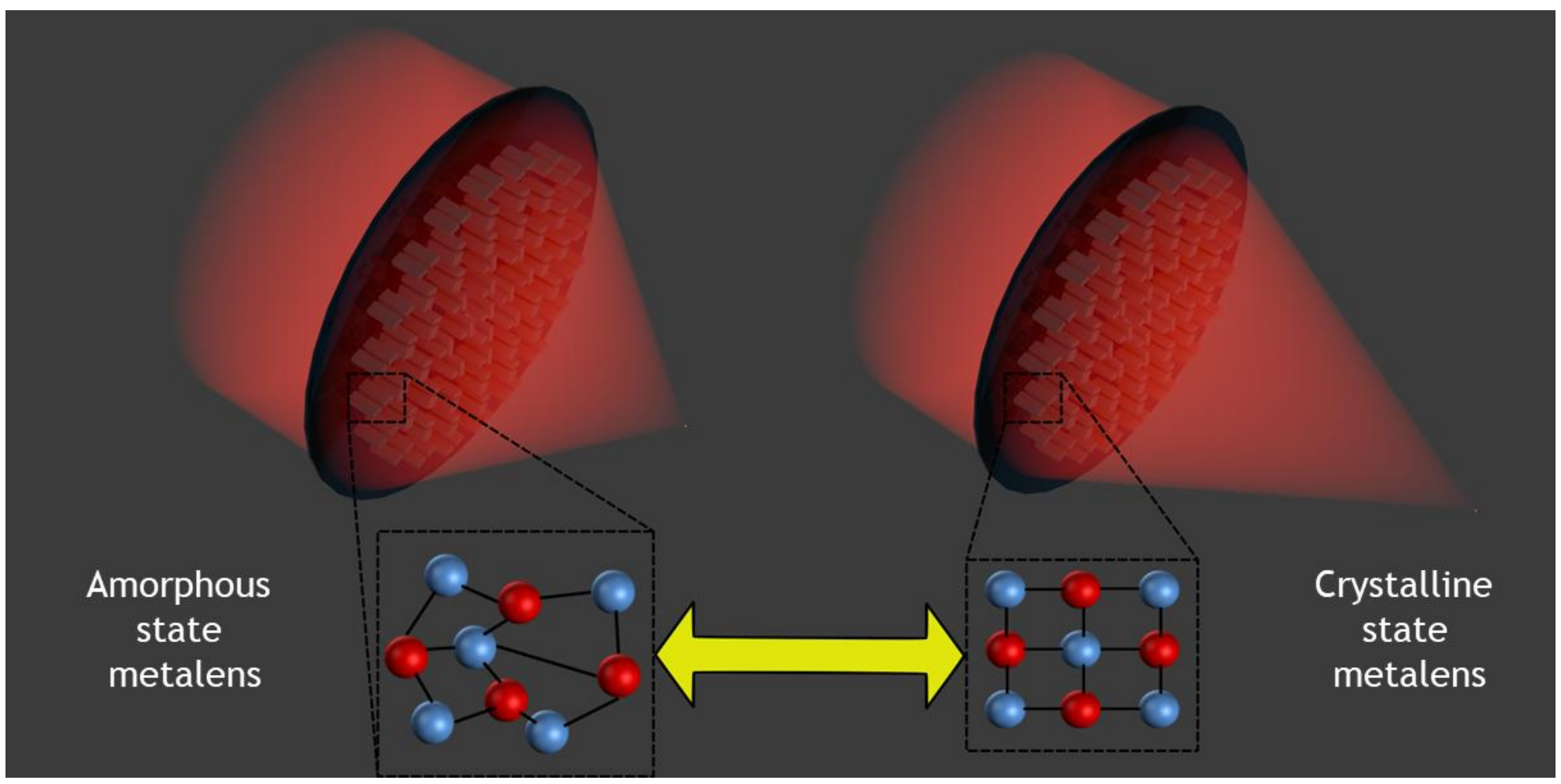

**Fig. 1. Conceptual illustration of the reconfigurable phase-change varifocal metalens.** Schematic depiction of a metalens based on a phase-change chalcogenide platform ($Sb_2S_3$) operating at the edge of the visible spectrum. The device switches between two distinct focal lengths by reversibly transforming the material between amorphous (left) and crystalline (right) states, thereby altering the effective refractive index and the imparted phase profile. This non-volatile transition enables varifocal functionality without mechanical motion or continuous power consumption. Insets illustrate the representative atomic configurations in the amorphous and crystalline phases, highlighting the origin of the optical contrast. The metalens aperture is composed of an array of subwavelength meta-atoms whose collective response implements the required phase modulation in each state.

Realizing reconfigurable metalenses capable of large focal length tuning with relatively short response times while still possessing reasonably high focussing efficiency and diffraction-limited operation in the visible region is therefore a key challenge. Phase-change chalcogenides, with their actively tunable optical properties, provide a promising platform for such devices, enabling non-volatile reconfiguration with potentially fast switching speeds [32–36]. However, realizing viable designs closer to the visible part of the spectrum is not trivial owing to intrinsic material absorption, relatively smaller size of the meta-atoms, as well as the constraints on the phase map, focusing efficiency and diffraction-limited operation across the different material states of the metalens platform. In this work, using a combination of finite element computations and genetic algorithm based inverse design strategies, we propose a practically feasible, reconfigurable $Sb_2S_3$-based metalens showing near-diffraction-limited operation at the edge of the visible region with 33% tunability in the focal length, focusing efficiencies of around 40% and Strehl's ratio of 0.8 and 0.77 in the amorphous and crystalline states, respectively, as shown in the concept schematic in Fig. 1.

## 2. Design of the varifocal metalens

### 2.1. General design considerations

We used a thin film of the phase change chalcogenide $Sb_2S_3$ as the metalens platform. Owing to its low loss and high refractive index values, it is particularly attractive for applications in the visible region [37,38]. We designed the metalens to have a central operational wavelength of 795 nm. The choice of wavelength was dictated by the fact that it is at the D1 transition of $^{87}$Rb atom, making it of particular interest in atomic, molecular and optical experiments. The refractive index of amorphous and crystalline states of $Sb_2S_3$ was adopted from Dong et al.,[39] (see Supplement 1, Fig. S1). We designed the meta-atom by employing the geometric phase modulation-based approach wherein each meta-atom was composed of two rectangular nanopillars for reasons elucidated later. The meta-atom geometry and arrangement were dictated by the constraint on the phase profile which must be parabolic for a convergent lens i.e., $\phi(x,y) = -\frac{2\pi}{\lambda}\left(\sqrt{x^2+y^2+f^2}-f\right)$. For practical reasons, we performed a 4-level discretization of the phase map. Fig. 2(a) and Fig. 2(b) show the discretized phase maps in the amorphous and crystalline states respectively. Endowing metalens with reconfigurability entails switching between two or more such independent phase maps via some tuning mechanism. A 4-level discretized phase map would require 16 different meta-atom geometries to switch between two different focal lengths in the amorphous and crystalline states. This implies that if we have some geometry that imparts an optical phase $\phi$ in the amorphous state, it should impart $\phi + \Delta\phi \; \forall \; \Delta\phi \in \{0, \pi/2, \pi, 3\pi/2, 2\pi\}$ in the crystalline state. A primary constraint to enable varifocal operation was to design four meta-atom geometries which have $\Delta\phi = 0, \pi/2, \pi$ and $3\pi/2, 2\pi$ between the amorphous and crystalline states with maximum transmission. Once these geometries were identified, they can be rotated to impart the required geometric phase in both the amorphous and crystalline states.

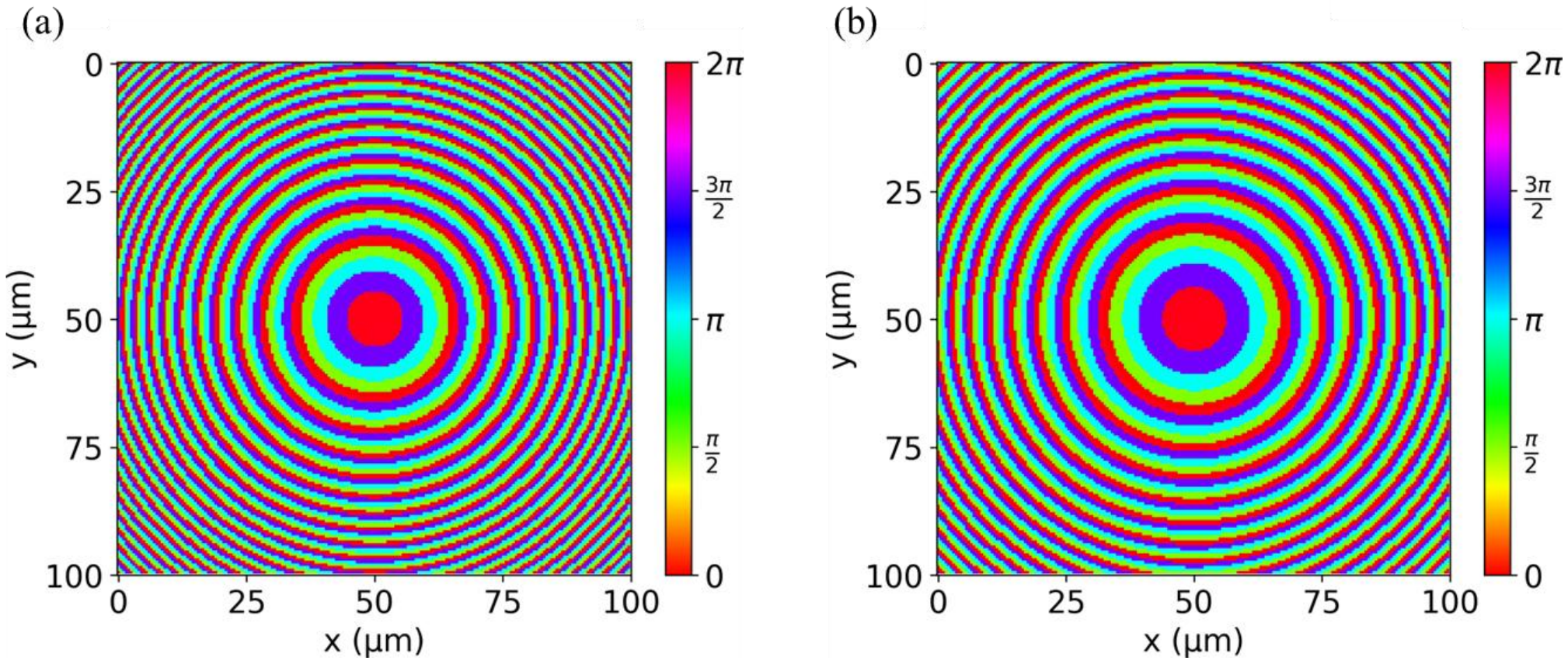

**Fig. 2. Phase profiles of the reconfigurable metalens in the two material states.** Spatial phase distributions corresponding to the discretized parabolic phase profiles used to realize focusing at 795 nm in the (a) amorphous and (b) crystalline states of the $Sb_2S_3$ metalens. The continuous ideal phase profile was quantized into four discrete phase levels to enable implementation using a finite meta-atom library. The colour scale denotes the phase accumulated across the aperture, spanning from 0 to $2\pi$. These discretized phase maps are to be mapped to physically realizable meta-atom geometries while preserving the required relative phase relationships between the amorphous and crystalline states.

### 2.2. Optimization via inverse design

The initial design employed a single $Sb_2S_3$ nanopillar as the meta-atom on a glass substrate, and finite-element simulations (see Supplement 1, section S1) were performed to evaluate the geometric phase imparted upon rotation of the meta-atom. These results indicated that the required set of inter-state phase differences, $\Delta\phi = 0$ or $2\pi$, $\pi/2$, $\pi$, $3\pi/2$ between the amorphous and crystalline states could not be achieved using a single nanopillar geometry. This limitation arises from the restricted anisotropy and phase dispersion accessible from a single dielectric scatterer, which constrains the independent control of phase response across the different material states. To overcome this limitation, we adopted a paired rectangular nanopillar configuration, which creates additional geometric degrees of freedom and enhances the effective anisotropy. This expanded design space enables more flexible control over the phase response in both amorphous and crystalline states, thereby allowing the required inter-state phase differences to be more readily satisfied [24]. The structural parameters including lengths and widths ($l_1$, $w_1$, $l_2$, $w_2$) of the nanopillars as well as the height and the periodicity were optimized to obtain geometries with high transmission while maintaining a particular phase difference imparted between the amorphous and crystalline states. Since searching for the relevant geometries through brute force techniques proved ineffective, we employed an inverse design optimization approach based on genetic algorithm. To reduce the computational complexity, we fixed the periodicity to 500 nm, the meta-atom height to 600 nm and the gap between the two nanopillars to 60 nm. A genetic algorithm (GA) based approach typically generates a library of meta-atoms by optimizing multiple objective functions while at the same time also satisfying constraints [40]. GA is a population-based, evolutionary optimization algorithm inspired by the process of natural selection. In each generation, a population of candidate solutions was evolved through biologically inspired operations such as selection, crossover, and mutation. Candidate designs were evaluated based on two objective functions, which quantify the transmission in both amorphous and crystalline states.

$$Obj_1 = 1 - T_{am} \tag{1}$$

$$Obj_2 = 1 - T_{cr} \tag{2}$$

where $T_{am}$ and $T_{cr}$ are the transmittance of the meta-atom in the amorphous and crystalline states, respectively. GA implementations usually follow the convention that high-performing individuals have a lower objective function. A constraint was also imposed to ensure that the phase difference between the amorphous and crystalline states is one of four values allowed for a four-level discretized phase map

$$\Delta\phi \in 0, \frac{\pi}{2}, \pi, \frac{3\pi}{2}, 2\pi \tag{3}$$

High-performing candidates were identified using a non-dominated sorting framework within a multi-objective genetic algorithm. Parent geometries were selected via tournament selection, and successive generations were produced through crossover and mutation of geometric parameters. By appropriately choosing the population size and convergence criteria, the algorithm yields a pareto-optimal set of meta-atoms, i.e., geometries for which no objective function can be improved without degrading another. The optimized meta-atom geometries were rotated to impart a geometric phase, enabling full $2\pi$ phase coverage in both amorphous and crystalline states.

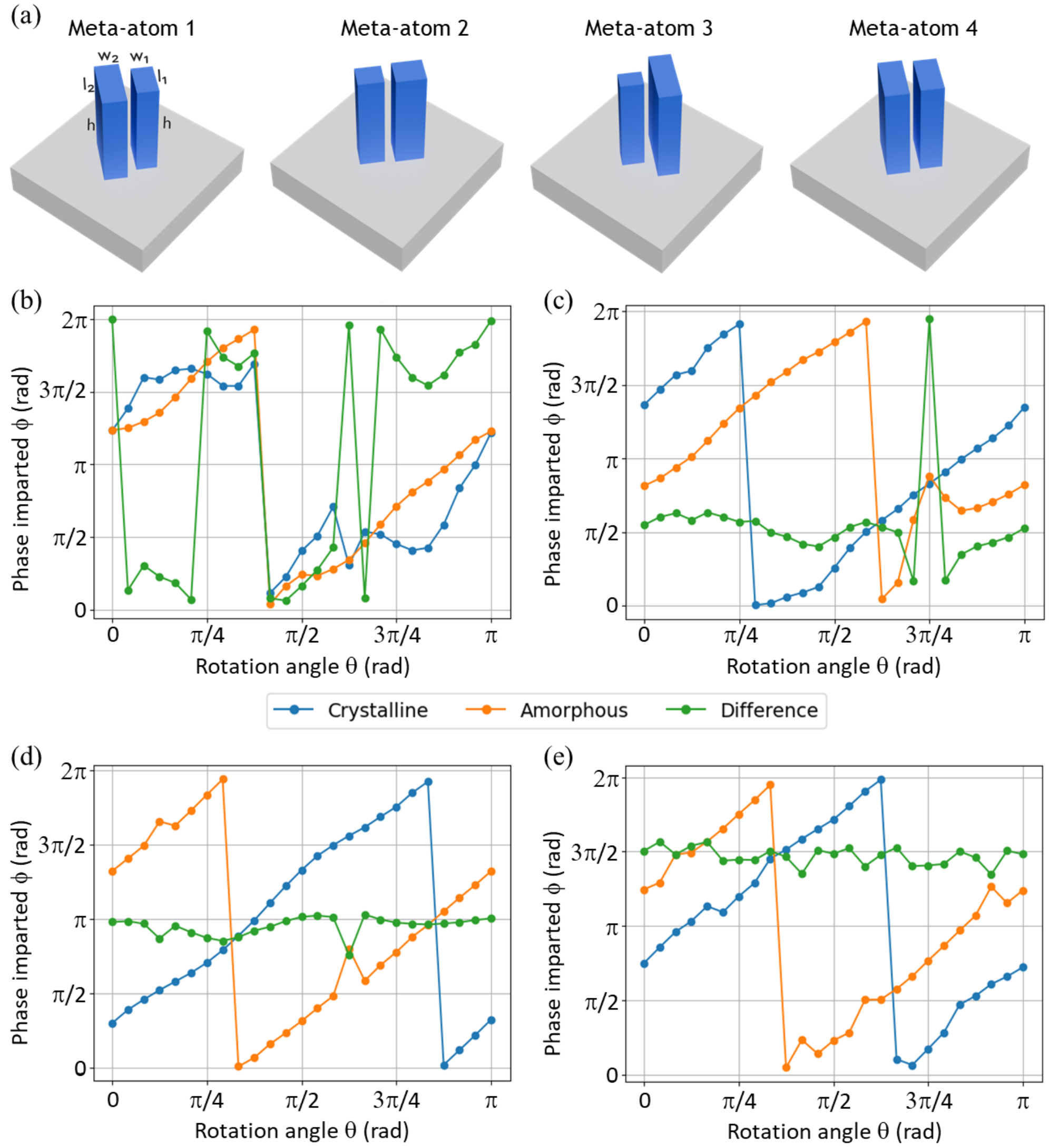

**Fig. 3. Inverse-designed optimal meta-atoms.** (a) Schematic representation of the four optimized meta-atom geometries, each comprising a pair of rectangular nanopillars with dimensions $l_1, l_2$(lengths) and $w_1, w_2$(widths), placed on a glass substrate. The height h of the nanopillars and the gap between them were fixed at 600 nm and 60 nm, respectively. The use of paired nanopillars provides additional geometric degrees of freedom required to satisfy multi-state phase constraints. Phase imparted as a function of rotation angle $\theta$ for (b) meta-atom 1, (c) meta-atom 2, (d) meta-atom 3 and (e) meta-atom 4, in the amorphous (orange) and crystalline (blue) states, along with the corresponding relative phase difference (green). Rotation of the meta-atoms enables geometric phase modulation, yielding full $2\pi$ phase coverage in both material states. The designs were optimized to realize prescribed relative phase offsets ($\phi_{cryst} - \phi_{amorp}$ = 0, $\pi/2$, $\pi$, $3\pi/2$) between the amorphous and crystalline states.

## 3. Optimized meta-atoms and metalens performance

Fig. 3(a) illustrates representative meta-atom geometries obtained through this optimization process. Section S1 (Figure S2) in Supplement 1 provides more information on the finite element method computation while Table S1 lists the geometrical parameters of the different meta-atoms. Figs. 3(b), (c), (d), and (e) show the geometric phase as a function of rotation angle for meta-atoms 1, 2, 3 and 4, respectively, in both amorphous and crystalline states, along with the corresponding relative phase difference. Rotation enables full $2\pi$ phase coverage while preserving a constant inter-state phase offset, specifically $\Delta\phi = \phi_{\text{cryst}} - \phi_{\text{amorp}} = 0$ or $2\pi$, $\pi/2$, $\pi$, $3\pi/2$ for meta-atoms 1, 2, 3, and 4, respectively. Minor deviations from the target phase offsets are observed across the rotation range, with a more noticeable deviation for meta-atom 2 near $\theta = 3\pi/4$. However, this localized deviation does not significantly impact overall device performance, as evidenced by the near-diffraction-limited Strehl ratios (~0.8) obtained for both material states as discussed later, indicating robustness of the design against such non-idealities.

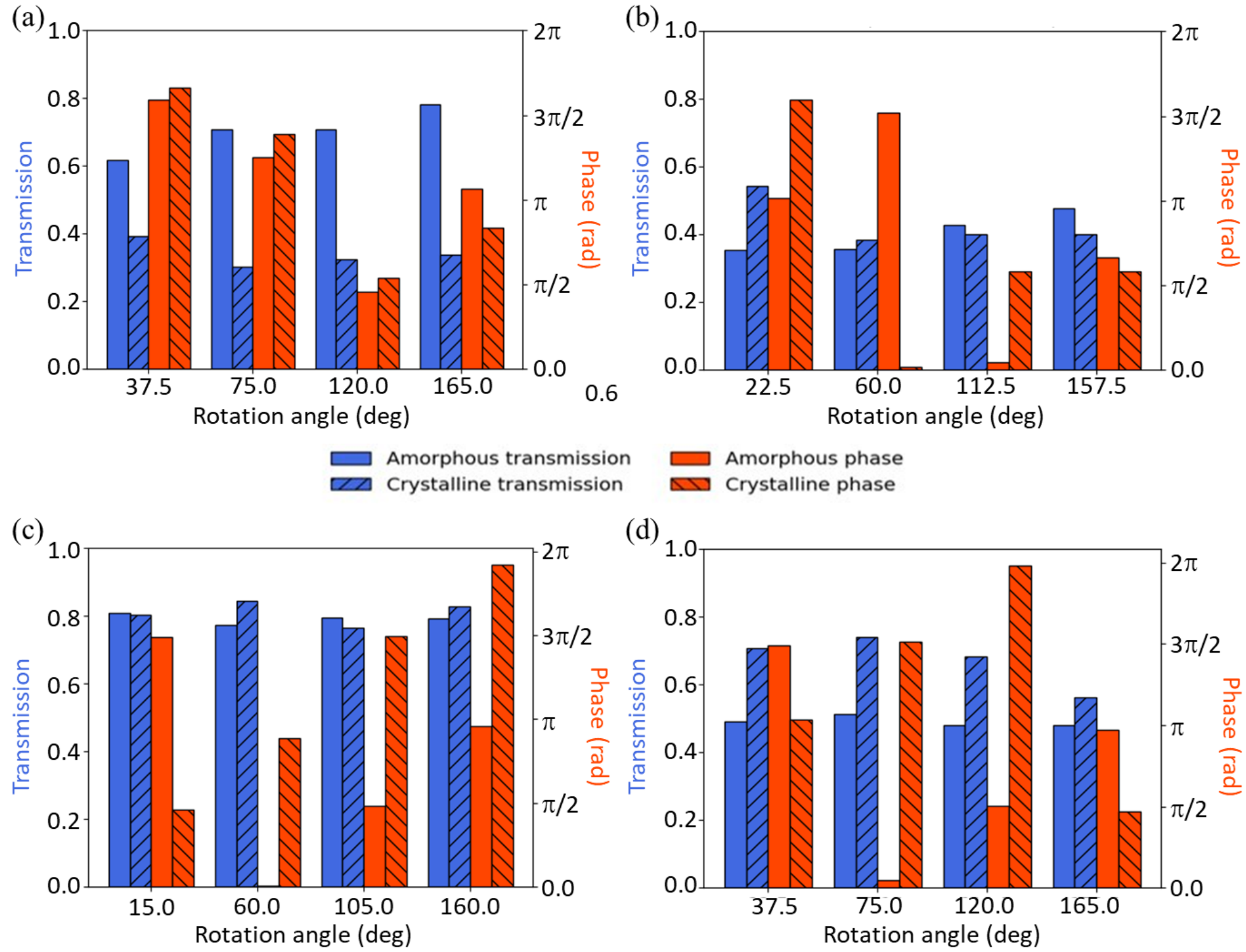

**Fig. 4. Optical response of the optimized meta-atoms in amorphous and crystalline states.** Transmission (blue bars, left axis) and phase (orange bars, right axis) responses of the four inverse-designed meta-atom geometries as a function of rotation angle for (a) meta-atom 1, (b) meta-atom 2, (c) meta-atom 3, and (d) meta-atom 4. Solid and hatched bars correspond to amorphous and crystalline states, respectively. The rotation of each meta-atom imparts geometric phase, enabling full $2\pi$ phase coverage while maintaining prescribed relative phase offsets between the two material states. The selected geometries achieve the target phase differences ($0, \pi/2, \pi, 3\pi/2$) with reasonably high transmission across both phases, forming the basis for constructing the discretized phase profiles of the reconfigurable metalens.

In addition to phase control, having reasonably good transmission in both material states is a critical requirement. Figs. 4(a)-4(d) collectively present the transmission coefficients and phase responses (evaluated as the magnitude and phase of the S21 parameter, see Supplement 1 for details) for the full set of 16 optimized meta-atoms, highlighting the trade-off between phase coverage, inter-state phase contrast, and transmission efficiency. While the transmission coefficient values vary across different meta-atoms depending on the relative phase difference they were designed to provide, all the meta-atoms show more than 25% transmission, with meta-atom 3 yielding close to 80% transmission irrespective of the rotation angle. The key metric in terms of transmission is that for the overall metalens obtained by putting together these meta-atoms as discussed below.

The generated meta-atom library was used to construct a reconfigurable metalens comprising a 200 × 200 array of unit cells with a period of 500 nm, corresponding to an overall aperture of 100 μm × 100 μm. The metalens was designed to achieve focal lengths of 150 μm and 200 μm in the amorphous and crystalline states, respectively. The ideal continuous phase profile was discretized into four phase levels by mapping each target phase value to the nearest available level and assigning the corresponding meta-atom from the library. The performance of the metalens constructed from the generated meta-atom library was evaluated through full-device simulations. Owing to the prohibitive computational cost of large-scale finite-element simulations, the intensity distribution at the focus was computed using the angular spectrum method as detailed in Section S2 of Supplement 1 [24,41]. In these computations, a normally incident plane wave with total power normalized to 1 W over the metalens aperture was assumed. Fig. 5(a) and Fig. 5(b) present the resulting axial (z-axis) intensity distributions for the metalens in the amorphous and crystalline states, respectively, while Fig. 5(c) and Fig. 5(d) show the corresponding transverse intensity profiles at the focal plane. The computed profiles exhibit the expected focusing behaviour, with clear and well-defined focal spots. Importantly, the focal length is accurately tuned between the designed values in the amorphous and crystalline states, confirming the effectiveness of the proposed meta-atom library and phase-engineering approach.

The efficacy of the metalens was then evaluated using two key metrics: Strehl ratio and focusing efficiency. Strehl ratio serves as an indicator of diffraction-limited performance and is defined as the ratio of the on-axis intensity of the realized metalens to that of a metalens with the ideal phase profile and is mathematically defined as [42]:

$$S = \frac{I_{ML}(0,0)}{I_{ideal}(0,0)} = \left|\langle e^{i\delta} \rangle\right| \tag{4}$$

where δ is the deviation of the real phase map from the ideal phase map. For a lens to be regarded as diffracted limited, the Strehl ratio must be greater than 0.8. The optimized reconfigurable metalens design yielded Strehl ratio of 0.80 in the amorphous state and 0.77 in the crystalline state, alluding to performance very close to the diffraction-limited regime. Notably, the slight degradation in the crystalline state can be attributed to increased phase errors arising from the combined effects of coarse phase discretization as well as localized deviations in the relative phase difference $\Delta\phi$ for one or two of the 16 meta-atom geometries . These results highlight that even with a limited four-level phase implementation, the inverse-designed meta-atom library enables sufficiently accurate phase reconstruction to approach diffraction-limited performance. Further improvements in Strehl ratio are expected with finer phase discretization or expanded meta-atom libraries, albeit at the cost of increased design complexity and computational effort.

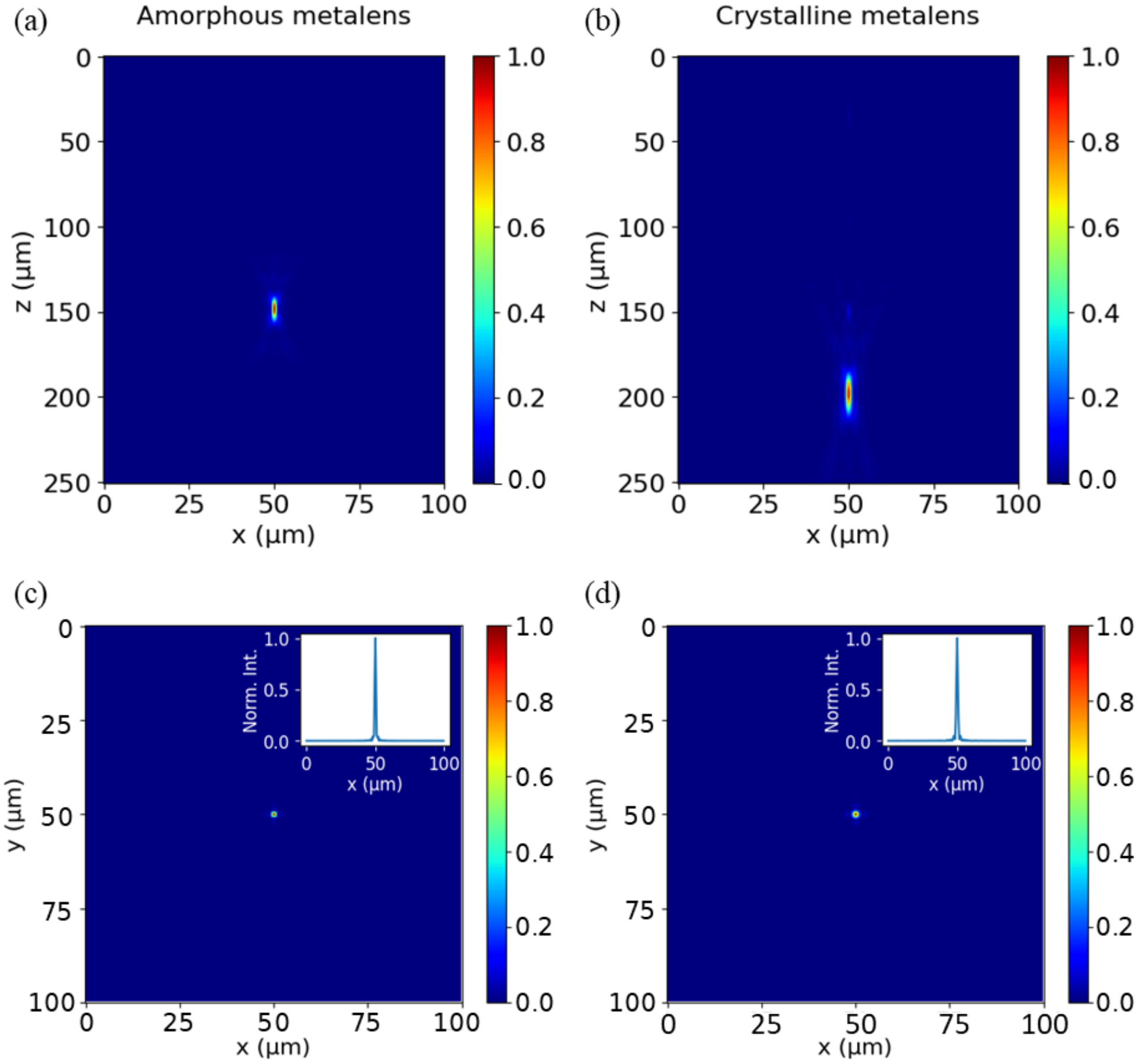

**Fig. 5. Simulated intensity distributions demonstrating varifocal operation of the reconfigurable metalens.** Normalized intensity profiles computed using the angular spectrum method for the $Sb_2S_3$ metalens under normal plane-wave illumination. (a,b) Axial (x–z) intensity maps showing the shift in focal position from $z \approx 150\,\mu$m in the amorphous state to $z \approx 200\,\mu$m in the crystalline state, confirming focal-length tunability via phase transition. (c,d) Corresponding transverse (x–y) intensity distributions at the respective focal planes, exhibiting well-defined, near-diffraction-limited focal spots. Insets show line cuts of the normalized intensity along the x-axis, highlighting the focal spot profile.

The second metric, focusing efficiency, is a measure of how well the metalens focuses light. This was computed by simulating the metalens operation via the beam propagation method which propagates wavefront starting from the complex electric field at the source plane to a destination plane through Fourier optics techniques (see Supplement 1, Section S2). The focusing efficiency was computed as:

$$\eta = \frac{P_{focus}}{P_{total}} \tag{5}$$

where $P_{focus}$ is the power contained within the FWHM of the intensity profile at the focus, and $P_{total}$ is the total power incident on the metalens, which is normalized to 1W. The focusing efficiency of the proposed metalens was calculated to be 43.7% in the amorphous phase and 38.3% in the crystalline phase. The linewidth of the focal spot (FWHM) was 3.20 μm for the amorphous state metalens and 4.12 μm for the crystalline state metalens calculated from a Gaussian fit of the line cut of the transverse focal spot (insets in the Figs. 5(c) and 5(d). The design proposed here can be integrated with an ITO-based microheater to facilitate electrical switching with rates in the range of 1 – 100 MHz [43]. While the focal-length tuning between 150 μm and 200 μm was chosen as a proof of principle, the achievable tuning range can, in principle, be extended by appropriately increasing the metalens aperture. However, such scaling must be approached with care to ensure the validity of the locally periodic approximation. As the phase profile becomes more rapidly varying with larger tuning ranges, this approximation may break down, necessitating full-wave simulations of the entire metalens, which could be computationally more intensive. We also note that focusing efficiency can potentially be improved by performing a more rigorous search for meta-atoms with higher transmission in both phases which is also computationally more demanding and hence, outside the scope of the present work. We also investigated the possibility of broadband operation by calculating the Strehl ratio and focusing efficiencies for a bandwidth surrounding the central wavelength of 795 nm. Considering that group delay compensation was not included as one of the optimization constraints, both the performance parameters showed a high degree of sensitivity to a change in wavelength. In particular, within a ±10 nm bandwidth, the Strehl ratio dropped to 0.6 for the amorphous metalens and 0.4 for the crystalline metalens while the focusing efficiency decreased to 26% and 11%, respectively.

## 4. Discussion

The metalens design presented in this work is unique in several respects. First, the device is realized entirely within the active phase-change medium, enabling maximum switching speeds ~ 100 MHz. Second, reconfiguration can be achieved through a global voltage pulse, eliminating the need for individually addressing groups of meta-atoms and thereby significantly simplifying device architecture and control. Third, the metalens operates near the visible spectrum while maintaining reasonable focusing efficiency and Strehl ratio, despite the increased material absorption typically encountered at shorter wavelengths. Finally, the design is compatible with standard nanofabrication processes, supporting practical scalability and integration. While a few $Sb_2S_3$-based varifocal metalens designs have been reported at telecommunication wavelengths, these approaches rely on meta-atom geometries and material tuning strategies that are not readily amenable to practical implementation [34,35]. In contrast, the present design adopts a physically realizable geometry and a robust switching mechanism, enabling reliable operation without imposing prohibitive fabrication or material constraints. Although the proposed metalens does not support continuous focal-length tuning, the ability to switch between two independently optimized phase profiles enables larger focal-length modulation than previously reported designs.

Importantly, this work establishes a practical and scalable pathway for implementing non-volatile, reconfigurable metalenses in challenging spectral regimes. The focusing efficiencies (~43.7% and ~38.3% in the amorphous and crystalline states, respectively) may appear modest in isolation; however, for metalenses realized entirely within an active phase-change chalcogenide medium, these values represent, to the best of our knowledge, among the highest reported. Future extensions could enable quasi-continuous tuning through partial crystallization of phase-change materials, as demonstrated for GSST at 1550 nm by Zhang et al. [44]. In addition, further refinement of the inverse design framework is expected to yield improved Strehl ratios and enhanced broadband performance, opening the door to high-efficiency, multifunctional tunable metasurfaces around the visible spectrum.

## 5. Conclusion

In summary, we have proposed a viable platform for active varifocal metalenses based on the amorphous-to-crystalline phase transition of $Sb_2S_3$ meta-atoms, enabled through a combination of finite-element simulations and genetic algorithm-based inverse design. The optimized device comprises four distinct meta-atom geometries based on paired rectangular nanopillar structures, achieving a focal length tunability of 33%, with focusing efficiencies of 43.7% and 38.3% in the amorphous and crystalline states, respectively. The metalens further demonstrates near-diffraction-limited performance, with Strehl ratios of 0.80 and 0.77 in the two material states. Importantly, this work establishes a physically realizable and scalable approach to non-volatile, reconfigurable metalenses operating near the visible spectrum, addressing key challenges in phase control, efficiency, and fabrication. The proposed architecture is well-suited for applications requiring dynamic focal-length tuning with fast switching speeds in the range of 1–100 MHz, and provides a promising pathway toward compact, high-performance adaptive photonic systems.

**Funding:** Department of Atomic Energy, Government of India, under Project Identification No. RTI 40007; Anusandhan National Research Foundation (ANRF) under projects CRG/2023/006217 and ANRF/ARG/2025/007930/PS.

**Conflict of Interest:** The authors have no conflict of interest to disclose.

**Data availability statement:** The data that supports the findings of this study are available from the corresponding author upon reasonable request.

**Supplemental document:** See Supplement 1 for supporting content.

## Supplemental Document

### S1. Discretized phase maps and Meta-atom library

This section gives additional information on the different meta-atom geometries obtained via inverse design strategies.

Fig. S1 shows the complex refractive index dispersion of $Sb_2S_3$ film in the amorphous and crystalline states. This was obtained by adopting the data from Dong et al and modelling it with a Tauc-Lorentz oscillator [39].

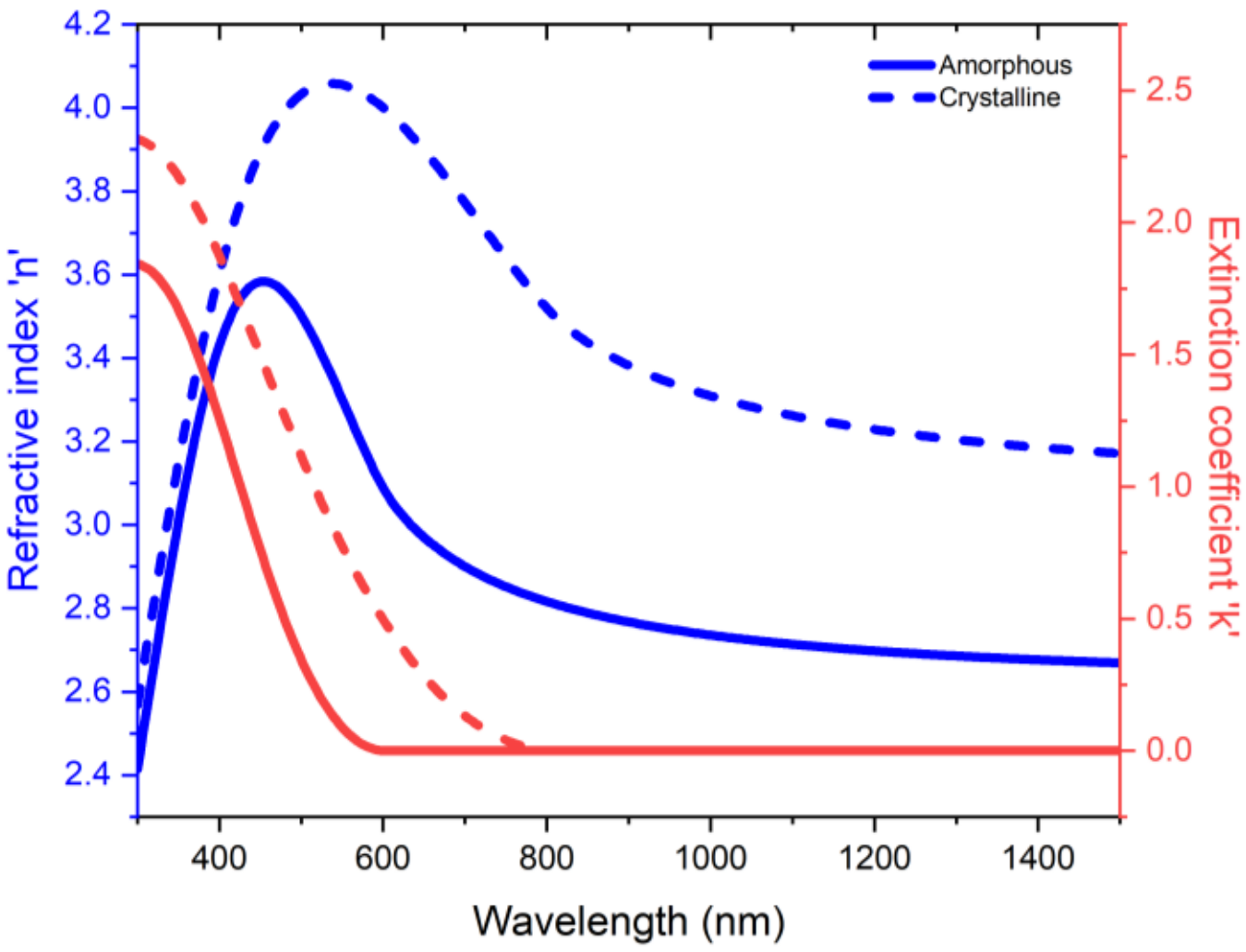

**Fig. S1:** Refractive index and extinction coefficient of the $Sb_2S_3$ film in the amorphous and crystalline states.

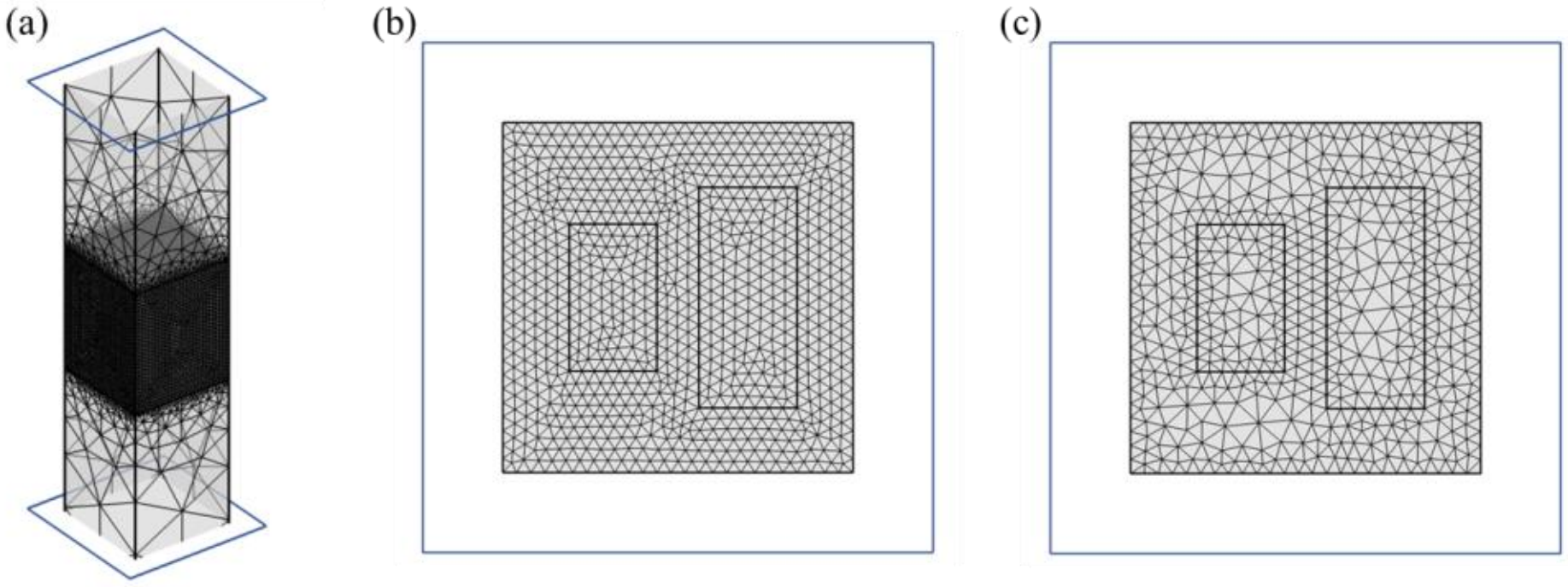

**Fig.S2:** (a) 3D view of the finite element simulation geometry with clipping planes enabled on the air (top) and glass substrate (bottom) sides for clarity showing the meshing methodology. Top-down (xy plane) view of the user-defined meshing in the meta-atom unit cell with (b) finest meshing, and (c) slightly coarser mesh employed to make the computation less intensive.

Optimization of the meta-atom geometry was performed via finite element simulations using the RF module in COMSOL Multiphysics®. Fig. S2 shows the simulation geometry and the user-defined meshing used in the simulations. We employed a very fine meshing for the meta-atom unit cell with the largest meshing element being 3 times smaller than the smallest dimension in the meta-atom geometry (10 – 20 nm) as shown in Fig. S2(b). To optimize the computational resources, we resorted to a slightly coarser mesh with the size of the largest element being 30 nm (Fig. S2(c)) after having verified that the physics and the results are captured accurately when moving from the very fine setting to the slightly coarser one. To calculate the transmittance and phase acquired due to the geometric phase design, each meta-atom geometry was excited with left circular polarized light from a port placed at the bottom of the simulation domain, and the transmission (amplitude and phase of S21 parameter) was computed by means of a port at the top of the simulation domain (on the air side). Periodic boundary conditions were employed along the horizontal ( x and y axes).

.

**Table S1** below lists the geometrical parameters and optical properties of the 16 different meta-atoms:

| **S.No.** | **label** | **$l_1$ (nm)** | **$w_1$ (nm)** | **$l_2$ (nm)** | **$w_2$ (nm)** | **θ (rad)** | **$T_{cry}$** | **$T_{amo}$** | **$\varphi_{cry}$ (rad)** | **$\varphi_{amo}$ (rad)** |
|---|---|---|---|---|---|---|---|---|---|---|
| 1 | MA 1 | 125 | 210 | 140 | 315 | 1.309 | 0.2639 | 0.8066 | 0.3623 | 0.1165 |
| 2 | MA 4 | 130 | 250 | 135 | 280 | 1.9635 | 0.6824 | 0.4794 | 6.2305 | 1.5878 |
| 3 | MA 3 | 130 | 345 | 135 | 105 | 2.618 | 0.8274 | 0.7921 | 6.0372 | 3.0129 |
| 4 | MA 2 | 175 | 160 | 150 | 135 | 1.0472 | 0.3844 | 0.3578 | 0.0585 | 4.7756 |
| 5 | MA 2 | 175 | 160 | 150 | 135 | 1.9635 | 0.4012 | 0.4275 | 1.8288 | 0.1453 |
| 6 | MA 1 | 125 | 210 | 140 | 315 | 2.0944 | 0.3255 | 0.7076 | 1.69 | 1.4411 |
| 7 | MA 4 | 130 | 250 | 135 | 280 | 2.618 | 0.5631 | 0.4808 | 1.4816 | 3.0468 |
| 8 | MA 3 | 130 | 345 | 135 | 105 | 0.2618 | 0.8034 | 0.8103 | 1.4417 | 4.6842 |
| 9 | MA 3 | 130 | 345 | 135 | 105 | 1.0472 | 0.8437 | 0.7744 | 2.7881 | 0.0311 |
| 10 | MA 2 | 175 | 160 | 150 | 135 | 2.7489 | 0.358 | 0.4774 | 3.3599 | 2.0883 |
| 11 | MA 1 | 125 | 210 | 140 | 315 | 2.8798 | 0.3369 | 0.7808 | 2.6287 | 3.345 |
| 12 | MA 4 | 130 | 250 | 135 | 280 | 0.3927 | 0.7082 | 0.3927 | 3.2426 | 4.6915 |
| 13 | MA 4 | 130 | 250 | 135 | 280 | 1.1718 | 0.7406 | 0.5144 | 4.7538 | 0.1512 |
| 14 | MA 3 | 130 | 345 | 135 | 105 | 1.8326 | 0.7656 | 0.7965 | 4.6972 | 1.5163 |
| 15 | MA 2 | 175 | 160 | 150 | 135 | 0.3927 | 0.5421 | 0.3548 | 5.0151 | 3.1831 |
| 16 | MA 1 | 125 | 210 | 140 | 315 | 0.6545 | 0.3932 | 0.6186 | 5.2148 | 4.9939 |

**Table S1**: Properties of the optimized 16 different meta-atoms including their dimensions, rotation angle as well as the transmission and phase in the amorphous and crystalline states.

## S2. Computation of the intensity profile of the metalens

The angular spectrum method was used to propagate a wavefront starting from the complex electric field at the source plane to a destination plane through Fourier optics techniques. We used the matrix of S-parameters generated from the COMSOL Multiphysics® model (detailed in section S1) to compute the electric field $f(x, y)$ just after the metalens:

$$f(x, y) = S(x, y) \cdot E_{in}(x, y)$$

where $S(x, y)$ corresponds to the matrix of $S_{21}$ parameters of meta-atoms arranged according to the discretized phase map $\phi_{dis}(x, y)$, that is $S(x, y) = S_{21}$ where $arg(S_{21}) \approx \phi_{dis}(x, y)$. The matrix $S(x, y)$ represents how the metalens manipulates the incident wavefront. This implies that now the function $f(x,y)$ represents the electric field just after the metalens. Now, we take a Fourier transform of this function

$$F\left(k_x, k_y, 0\right) = \iint_{-\infty}^{\infty} f(x, y) e^{-i\left(k_x x + k_y y\right)} dx\, dy$$

Assuming the waves propagate in a linear medium, we can compute this Fourier transform at some location z by using the transfer function of the medium

$$F\left(k_x, k_y, k_z\right) = \mathcal{H}(k_z; z) F\left(k_x, k_y, 0\right)$$

where $\mathcal{H}(k_z; z) = e^{\pm i k_z z}$ is the transfer function of air.

To retrieve the electric field at z, we simply perform an inverse Fourier transform:

$$f(x, y, z) = \iint_{-\infty}^{\infty} F\left(k_x, k_y, k_z\right) e^{i\left(k_x x + k_y y\right)} dk_x\, dk_y$$

$$\Rightarrow f(x, y, z) = \iint_{-\infty}^{\infty} F\left(k_x, k_y, 0\right) e^{i\left(k_x x + k_y y\right)} e^{\pm i k_z z} dk_x\, dk_y$$

Here

$$k_z = \sqrt{k^2 - k_x^2 - k_y^2}\,;\ k = {}^{2\pi}/_{\lambda}$$

$$k_x = k \sin\theta_x\,, k_y = k \sin\theta_y$$

with $\theta_x = \sin^{-1}(\lambda \nu_x)$ and $\theta_y = \sin^{-1}(\lambda \nu_y)$

where $\nu_x$ and $\nu_y$ are the spatial frequencies of the metalens.

At any point z beyond the metalens, the intensity profile of the resultant field can be computed as:

$$I(x, y, x) = |f(x, y, z)|^2$$

In practice, the intensity profile has to be discretized for numerical simulation. We chose a grid size of 200 x 200, which corresponds to one point per meta-atom.